\documentstyle[aaai98,epsf,rotate]{article} 

\title{\vspace{-0.35in}
Anytime Coalition Structure Generation with Worst Case Guarantees\thanks{\hspace{0.01in} Copyright 1998, American Association for Artificial Intelligence
(www.aaai.org). All rights reserved.  Supported by NSF CAREER award IRI-9703122 and NSF grant IRI-9610122.}}

\author{Tuomas Sandholm$^1$ Kate Larson$^2$ Martin Andersson$^3$ Onn Shehory$^4$ Fernando Tohm\'e$^5$ \vspace{-0.2in}
\AND
\normalsize $^{1,2,3,5}$ Washington University \vspace{-0.03in}\\
\normalsize Department of Computer Science \vspace{-0.03in}\\
\normalsize St. Louis MO 63130-4899 \vspace{-0.03in}\\
\normalsize \{sandholm,ksl2,mra,tohme\}@cs.wustl.edu
\And
\normalsize $^4$ Carnegie Mellon University \vspace{-0.03in}\\
\normalsize The Robotics Institute \vspace{-0.03in}\\
\normalsize Pittsburgh PA 15213-3890 \vspace{-0.03in}\\
\normalsize onn@cs.cmu.edu
}

\newtheorem{lemma}{Lemma}
\newtheorem{theorem}{Theorem}

\newtheorem{algorithm}{Algorithm}

\newenvironment{proof}{\noindent {\bf Proof. }}{${}_\Box$ \\}
\newenvironment{proofnobox}{\noindent {\bf Proof. }}{}

\begin{document}
\maketitle


\newcommand{\argmax}[1] {\arg \hspace{-0.04in} \max_{{}_{#1}}}
\newcommand{\argmin}[1] {\arg \hspace{-0.04in} \min_{{}_{#1}}}

\newcommand{\ttchoose}[2] {\left( \begin{array}{c} {#1} \\ {#2} \end{array} \right)}

\newcommand{\ttalts}[4] {\left\{ \begin{array}{ll} {#1}   & \mbox{#2} \\
                                                  {#3} & \mbox{#4}
                                 \end{array} \right. }
\newcommand{\threealts}[6] {\left\{ \begin{array}{ll} {#1}   & \mbox{#2}\\
                                                      {#3}   & \mbox{#4}\\
                                                      {#5}   & \mbox{#6}
                                 \end{array} \right. }
\newcommand{\sij}[2] {\left. \begin{array}{ll} {}   & \mbox{#1} \\
                                               {}   &  {}       \\
                                               {}   & \mbox{#2}
                                 \end{array}
                         \right| }

\begin{abstract}
Coalition formation is a key topic in multiagent systems.  One would
prefer a coalition structure that maximizes the sum of the values of
the coalitions, but often the number of coalition structures is too
large to allow exhaustive search for the optimal one.  But then, can
the coalition structure found via a partial search be guaranteed to be
within a bound from optimum?


We show that none of the previous coalition structure generation
algorithms can establish any bound because they search fewer nodes
than a threshold that we show necessary for establishing a bound.  We
present an algorithm that establishes a tight bound within this
minimal amount of search, and show that any other algorithm would have
to search strictly more.  The fraction of nodes needed to be searched
approaches zero as the number of agents grows.

If additional time remains, our anytime algorithm searches further,
and establishes a progressively lower tight bound.  Surprisingly, just
searching one more node drops the bound in half.  As desired, our
algorithm lowers the bound rapidly early on, and exhibits diminishing
returns to computation.  It also drastically outperforms its obvious
contenders.  Finally, we show how to distribute the desired search
across self-interested manipulative agents.



\end{abstract}


\section{Introduction}

Multiagent systems with self-interested agents are becoming
increasingly important.  One reason for this is the {\em technology
push} of a growing standardized communication
infrastructure---Internet, WWW, NII, EDI, KQML, FIPA, Concordia,
Voyager, Odyssey, Telescript, Java, {\em etc}---over which separately
designed agents belonging to different organizations can interact in
an open environment in real-time and safely carry out
transactions~\cite{Sandholm97:Safe}.
The second reason is strong {\em application pull} for computer
support for negotiation at the operative decision making level.  For
example, we are witnessing the advent of small transaction commerce on
the Internet for purchasing goods, information, and communication
bandwidth.
There is also an industrial trend toward virtual enterprises: dynamic
alliances of small, agile enterprises which together can take
advantage of economies of scale when available (e.g., respond to more
diverse orders than individual agents can), but do not suffer from
diseconomies of scale.

Multiagent technology facilitates the automated formation of such
dynamic coalitions at the operative decision making level.  This
automation can save labor time of human negotiators, but in addition,
other savings are possible because computational agents can be more
effective at finding beneficial short-term coalitions than humans are
in strategically and combinatorially complex settings.

This paper discusses coalition structure generation in settings where
there are too many coalition structures to enumerate and evaluate due
to, for example, costly or bounded computation and/or limited time.
Instead, agents have to select a subset of coalition structures on
which to focus their search.  We study which subset the agents should
focus on so that they are guaranteed to reach a coalition structure
that has quality within a bound from the quality of the optimal
coalition structure.


\section{Coalition formation setting}

In many domains, self-interested real world parties---e.g., companies
or individual people---can save costs by coordinating their activities
with other parties.  For example, when the planning activities are
automated, it can be useful to automate the coordination activities as
well.  This can be done via a negotiating software agent representing
each party.  Coalition formation includes three activities:
\begin{enumerate}

\item {\em Coalition structure generation}: formation of coalitions by
the agents such that agents within each coalition coordinate their
activities, but agents do not coordinate between coalitions.
Precisely, this means partitioning the set of agents into exhaustive
and disjoint coalitions.  This partition is called a {\em coalition
structure} ($CS$).

\item {\em Solving the optimization problem} of each coalition.  This
means pooling the tasks and resources of the agents in the coalition,
and solving this joint problem.    
The coalition's objective is to maximize monetary value: money
received from outside the system for accomplishing tasks minus the
cost of using resources.
%

\item {\em Dividing the value} of the generated solution among agents.

\end{enumerate}
These activities interact. For example, the coalition that an agent
wants to join depends on the portion of the value that the agent would
be allocated in each potential coalition.

This paper focuses on settings were the coalition structure generation
activity is resource-bounded: not all coalition structures can be
enumerated.  




\section{Our model of coalition structure generation}
\label{se:coalstruct_model}

Let $A$ be the set of agents, and $a = |A|$.  As is common
practice~\cite{Kahan84:Theories,Shehory95:Task,Shehory96:Kernel,Zlotkin:aaai94,Ketchpel:aaai94,Sandholm97:Coalition_incl95},
we study coalition formation in {\em characteristic function games}
(CFGs).  In such games, the value of each coalition $S$ is given by a
characteristic function $v_S$.\footnote{These coalition values $v_S$
may represent the quality of the optimal solution for each coalition's
optimization problem, or they may represent the best bounded-rational
value that a coalition can get given limited or costly computational
resources for solving the
problem~\cite{Sandholm97:Coalition_incl95}.}\footnote{In other words,
each coalition's value is independent of nonmembers' actions.
However, in general the value of a coalition may depend on nonmembers'
actions due to positive and negative externalities (interactions of
the agents' solutions).  Negative externalities between a coalition
and nonmembers are often caused by shared resources.  Once nonmembers
are using the resource to a certain extent, not enough of that
resource is available to agents in the coalition to carry out the
planned solution at the minimum cost.  Negative externalities can also
be caused by conflicting goals.  In satisfying their goals, nonmembers
may actually move the world further from the coalition's goal
state(s)~\cite{Rosenschein94:Rules}.  Positive externalities are often
caused by partially overlapping goals.  In satisfying their goals,
nonmembers may actually move the world closer to the coalition's goal
state(s).  From there the coalition can reach its goals less
expensively than it could have without the actions of nonmembers.
General settings with possible externalities can be modeled as {\em
normal form games} (NFGs).  CFGs are a strict subset of NFGs.
However, many real-world multiagent problems happen to be
CFGs~\cite{Sandholm97:Coalition_incl95}.}  We assume that $v_S$ is
bounded from below for each coalition $S$, i.e. no coalition's value
is infinitely negative.  We normalize the coalition values by
subtracting at least $\min_{S\subset A} v_S$ from all coalition values
$v_S$.\footnote{All of the claims of the paper are valid as long as
$v_S \geq 0$ for the coalitions that the algorithm {\em sees}:
coalitions not seen during the search may be arbitrarily bad.}  This
rescales the coalition values so that $v_S \geq 0$ for all coalitions
$S$.  This rescaled game is strategically equivalent to the original
game.

A coalition structure $CS$ is a partition of agents, $A$, into
coalitions.  Each agent belongs to exactly one coalition.  Some agents
may be alone in their coalitions.  We will call the set of all
coalition structures $M$.  For example, in a game with three agents,
there are 7 possible coalitions: $\{$1$\}$, $\{$2$\}$, $\{$3$\}$,
$\{$1,2$\}$, $\{$2,3$\}$, $\{$3,1$\}$, $\{$1,2,3$\}$ and 5 possible
coalition structures: $\{\{$1$\}$, $\{$2$\}$, $\{$3$\}\}$,
$\{\{$1$\}$, $\{$2,3$\}\}$, $\{\{$2$\}$, $\{$1,3$\}\}$, $\{\{$3$\}$,
$\{$1,2$\}\}$, $\{\{$1,2,3$\}\}$.

Usually the goal is to maximize the social welfare of the agents $A$
by finding a coalition structure
\begin{equation}
CS^* = \argmax{CS \in M} V(CS),
\end{equation}
where
\begin{equation}
V(CS) = \sum_{S \in CS} v_S
\label{eq:sumvss}
\end{equation}



The problem is that the number of coalition structures is large
($\Omega(a^{a/2})$), so not all coalition structures can be
enumerated---unless the number of agents is extremely small (below 15
or so in practice).\footnote{The exact number of coalition structures is
$\sum_{i=1}^a S(a, i)$, where $S(a, i) = i S(a-1, i) + S(a-1, i-1)$,
and $S(a,a)=S(a,1)=1$.}  Instead, we would like to search through a
subset $N \subseteq M$ of coalition structures, pick the best
coalition structure we have seen
\begin{equation}
CS^*_N = \argmax{CS \in N} V(CS),
\end{equation}
and be guaranteed that this coalition structure is within a bound from
optimal, i.e. that
\begin{equation}
k \geq \frac{V(CS^*)}{V(CS^*_N)}
\label{eq:k_def}
\end{equation}
is finite, and as small as possible.  We define $n_{min}$ to be the
smallest size of $N$ that allows us to establish such a bound $k$.


\begin{table}[htb]
{\footnotesize
\begin{center}
\begin{tabular}{|c|l|}
\hline
$A$ & The set of agents.\\
$a$ & The number of agents, i.e. $|A|$.\\
$S$ & Symbol for a coalition.\\
$CS$ & Symbol for a coalition structure.\\
$CS^*$ & Welfare maximizing coalition structure.\\
$M$ & The set of all possible coalition structures.\\
$m$ & $|M|$, total number of coalition structures.\\
$N$ & Coalition structures searched so far.\\
$n$ & $|N|$.\\
$n_{min}$ & Minimum $n$ that guarantees a bound $k$.\\
$CS^*_N$ & Welfare maximizing CS among ones seen.\\
$V(CS)$ & Value of coalition structure $CS$.\\
$k$ & Worst case bound on value, see Eq.~\ref{eq:k_def}.\\
\hline
\end{tabular}
\end{center}
\vspace{-0.1in}
\caption{\it Important symbols used in this paper.}
\label{ta:coalstruct_symbols}
}
\end{table}


\subsection{Lack of prior attention}

Coalition structure generation has not previously received much
attention.
Research has
focused~\cite{Kahan84:Theories,Zlotkin:aaai94} on superadditive games,
i.e. games where $v_{S \cup T} \geq v_S + v_T$ for all disjoint
coalitions $S,T \subseteq A$.  In such games, coalition structure
generation is trivial because the agents are best off by forming the
grand coalition where all agents operate together.  In other words, in
such games, $\{ A \}$ is a social welfare maximizing coalition
structure.

Superadditivity means that any pair of coalitions is best off by
merging into one.  Classically it is argued that almost all games are
superadditive because, at worst, the agents in a composite coalition
can use solutions that they had when they were in separate coalitions.

However, many games are not superadditive because there is some cost
to the coalition formation process itself.  For example, there might
be coordination overhead like communication costs, or possible
anti-trust penalties.  Similarly, solving the optimization problem of
a composite coalition may be more complex than solving the
optimization problems of component coalitions.  Therefore, under
costly computation, component coalitions may be better off by not
forming the composite coalition~\cite{Sandholm97:Coalition_incl95}.  Also, if
time is limited, the agents may not have time to carry out the
communications and computations required to coordinate effectively
within a composite coalition, so component coalitions may be more
advantageous.

In games that are not superadditive, some coalitions are best off
merging while others are not.  In such cases, the social welfare
maximizing coalition structure varies. This paper focuses on games
that are not superadditive (or if they are, this is not known in
advance).  In such settings, coalition structure generation is highly
nontrivial.

\section{Search graph for coalition structure generation}

\begin{figure}[hbt]
\epsfxsize=1.70in
\rotate[r]{
\epsffile{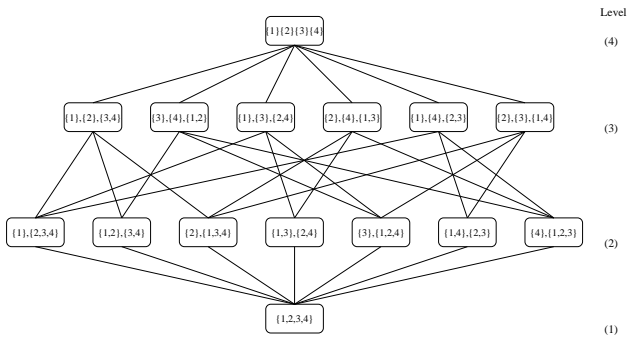}
}
\caption[Coalition structure graph.]{{\it Coalition structure graph
for a 4-agent game.  The nodes represent coalition structures. The
arcs represent mergers of two coalition when followed downward, and
splits of a coalition into two coalitions when followed upward.}}
\label{fi:coal_graph}
\end{figure}

Taking an outsider's view, the coalition structure generation process
can be viewed as search in a {\em coalition structure graph},
Figure~\ref{fi:coal_graph}.  Now, how should such a graph be searched
if there are too many nodes to search it completely?



\section{Minimal search to establish a bound}

This section discusses how a bound $k$ can be established while
searching as little of the graph as possible.

\begin{theorem}
To bound $k$, it suffices to search the lowest two levels of the
coalition structure graph (Figure~\ref{fi:coal_graph}).  With this
search, the bound $k=a$, and the number of nodes searched is
$n=2^{a-1}$.
\label{th:low2_sufficient}
\end{theorem}
\begin{proof}
To establish a bound, $v_S$ of each coalition $S$ has to be observed
(in some coalition structure).  The $a$-agent coalition can be
observed by visiting the bottom node.  The second lowest level has
coalition structures where exactly one subset of agents has split away
from the grand coalition.  Therefore, we see all subsets at this level
(except the grand coalition).  It follows that a search of the lowest
two levels sees all coalitions.

In general, $CS^*$ can include at most $a$ coalitions.  Therefore,
\begin{displaymath}
V(CS^*) \leq a \max_S v_S \leq a \max_{CS \in N} V(CS) = a V(CS^*_N).
\end{displaymath}
Now we can set $k=a\geq \frac{V(CS^*)}{V(CS^*_N)}$.

The number of coalition structures on the lowest level is 1.  The
number of coalitions on the second lowest level is $2^a -2$ (all
subsets of $A$, except the empty set and the grand coalition).  There
are two coalitions per coalition structure on this level, so there are
$\frac{2^a -2}{2}$ coalition structures at the second to lowest level.
So, there are $1 + \frac{2^a -2}{2} = 2^{a-1}$ coalition structures
(nodes) on the lowest two levels.
\end{proof}

\begin{theorem}
For the algorithm that searches the two lowest levels of the graph,
the bound $k=a$ is tight.
\label{th:tight}
\end{theorem}
\begin{proof}
We construct a worst case via which the bound is shown to be tight.
Choose $v_S = 1$ for all coalitions $S$ of size 1, and $v_S = 0$ for
the other coalitions.  Now, $CS^* = \{ \{1\}, \{2\}, ... , \{a\} \}$,
and $V(CS^*) = a$.  Then $CS^*_N = \{ \{1\}, \{2, ... , a\}
\}$.~\footnote{This is not unique because all coalition structures
where one agent has split off from the grand coalition have the same
value.}  Because $V(CS^*_N) = 1$, $\frac{V(CS^*)}{V(CS^*_N)} =
\frac{a}{1} =a$.
\end{proof}

\begin{theorem}
No other search algorithm (than the one that searches the bottom two
levels) can establish a bound $k$ while searching only $n = 2^{a-1}$
nodes or fewer.
\label{th:unique_alg}
\end{theorem}
\begin{proof}
In order to establish a bound $k$, $v_{S}$ of each coalition $S$ must
be observed.  The node on the bottom level of the graph must be
observed since it is the only node where the grand coalition appears.
Assume that the algorithm omits $m$ nodes on the second level.  Each
of the omitted nodes has $CS=\{P,Q\}$.  Since coalitions $P$ and $Q$
are never again in the same coalition structure, two extra nodes in
the graph have to be visited to observe $v_{P}$ and $v_{Q}$.  Assume
$m$ coalition structures $\{P_{1},Q_{1}\},\{P_{2},Q_{2}\},\ldots
,\{{P}_{m},Q_{m}\}$ are omitted.  Since for $i,j, i\not =j$, at least one
of the following is true, $P_{i}\cap P_{j}\not =\emptyset, P_{i} \cap
Q_{j} \not = \emptyset $, or $Q_{i}\cap Q_{j}\not = \emptyset $, at
least $m+1$ coalition structures must be visited to replace the $m$
coalition structure omitted.  Therefore, for the algorithm to
establish $k$, it must search $n>2^{a-1}$ nodes.
\end{proof}

So, $n_{min} = 2^{a-1}$, and this is uniquely established via a search
algorithm that visits the lowest two levels of the graph (order of
these visits does not matter).

\subsection{Positive interpretation}

\begin{figure}[hbt]
\vspace{-0.15in}
\epsfysize=2.5in
\epsfxsize=3.5in
\hspace*{\fill}
\hspace{-0.3in}
\epsffile{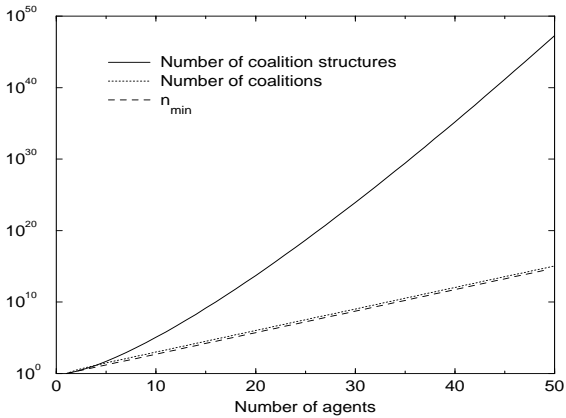}
\hspace*{\fill}
\vspace{-0.3in}
\caption[Number of coalition structures, coalitions, and coalition
structures needed to be searched.]{{\it Number of coalition
structures, coalitions, and coalition structures needed to be
searched.  We use a logarithmic scale on the value axis;
otherwise $n_{min}$ and the number of coalitions would be so small
compared to the number of coalition structures that their curves would
be indistinguishable from the category axis.}}
\label{fi:coals_vs_structs}
\end{figure}

Interpreted positively, our results (Theorem~\ref{th:low2_sufficient})
show that---somewhat unintuitively---a worst case bound from optimum
can be guaranteed without seeing all $CS$s.  Moreover, as the number
of agents grows, the fraction of coalition structures needed to be
searched approaches zero, i.e. $\frac{n_{min}}{m} \rightarrow 0$ as $a
\rightarrow \infty$.  This is because the algorithm needs to see only
$2^{a-1}$ coalition structures while the total number of coalition
structures is $\Omega(a^{a/2})$.  See Figure~\ref{fi:coals_vs_structs}.

%
%
%
%

\subsection{Interpretation as an impossibility result}

Interpreted negatively, our results
(Theorem~\ref{th:unique_alg}) show that exponentially
many ($2^{a-1}$) coalition structures have to be searched before a
bound can be established.  This may be prohibitively complex if the
number of agents is large---albeit significantly better than
attempting to enumerate all coalition structures.

Viewed as a general impossibility result, Theorem~\ref{th:unique_alg}
states that no algorithm for coalition structure generation can
establish a bound in general characteristic function games without
trying at least $2^{a-1}$ coalition structures.  This sheds light on
earlier algorithms.  Specifically, all prior coalition structure
generation algorithms for general characteristic function
games~\cite{Shehory96:Kernel,Ketchpel:aaai94}---which we know
of---fail to establish such a bound.  In other words, the coalition
structure that they find may be arbitrarily far from optimal.

\section{Lowering the bound with further search}

We have devised the following algorithm that will establish a bound in
the minimal amount of search, and then rapidly reduce the bound
further if there is time for more search.\footnote{If the domain
happens to be superadditive, the algorithm finds the optimal coalition
structure immediately.}

\noindent \rule{\columnwidth}{0.5mm}
\begin{algorithm}
\newcounter{coalstruct_alg1}
{\bf COALITION-STRUCTURE-SEARCH-1}
\vspace{-0.02in}
\begin{list}%
{\arabic{coalstruct_alg1}.}{\usecounter{coalstruct_alg1} \itemsep0.05in
  \itemindent0in \listparindent0in \labelwidth0.18in \leftmargin0.18in }


\item Search the {\bf bottom} two levels of the coalition structure graph.

\item Continue with a breadth-first search from the {\bf top} of the
graph as long as there is time left, or until the entire graph has
been searched (this occurs when this breadth-first search completes
level 3 of the graph, i.e. depth $a-3$).

\item Return the coalition structure that has the highest welfare
among those seen so far.

\end{list}
\label{al:coal_search}
\end{algorithm}
\vspace{-0.2in}
\noindent \rule{\columnwidth}{0.5mm}
\vspace{0.02in}

In the rest of this section, we analyze how this algorithm reduces the
worst case bound, $k$, as more of the graph is searched.  The analysis
is tricky because the elusive worst case ($CS^*$) moves around in the
graph for different searches, $N$.  We introduce the notation
$h=\lfloor\frac{a-l}{2}\rfloor +2$, which is used throughout this
section.
\begin{lemma}
\label{lem-pairing}
Assume that Algorithm~\ref{al:coal_search} has
just completed searching level $l$.  Then
\begin{enumerate}
\item If $a\equiv l \: (\mbox{mod } 2)$  coalitions of size $h$ will have been seen 
paired together with all coalitions of size $h-2$ or smaller.
\item If $a \not \equiv l \:(\mbox{mod } 2)$  coalitions of size $h$ will have been
seen paired together with all coalitions of size $h-1$ and smaller.
\end{enumerate}
\end{lemma}

\begin{proofnobox}
\begin{enumerate}
\item  At level $l$ the largest coalition in 
any coalition structure has size $a-l+1$.  Therefore, one of the
coalition structures at level $l$ is of the form $S_{1},S_{2},\ldots
,S_l$ where $|S_{i}|=1$ for $i<l$ and $|S_l|=a-l+1$.  Since
$a\equiv l \: (\mbox{mod }2)$, $h=\frac{a-l}{2}+2$.  Take coalition
$S_l$ and remove $h$ agents from it.  Call the new coalition formed
by the $h$ agents $S_l'$.  We will distribute the remaining
$\frac{a-l}{2}-1$ agents among the coalitions of size 1.  By doing
this we can enumerate all possible coalitions that can appear pairwise
with coalition $S'_l$ on level $l$.  For all $j=1,2,\ldots
,\frac{a-l}{2}-1$, place $\frac{a-1}{2}-j$ agents in coalition $S_{1}$
and call the new coalition $S_{1}^{j}$.  Redistribute the remaining
$j-1$ agents among coalitions $S_{2},\ldots ,S_{l-1}$.  For each $j$
we have listed a coalition structure containing both $S_l'$ and
$S^{j}_{1}$.  The largest of these $S_{1}^{j}$ has size
$\frac{a-l}{2}$, or $h-2$.

\item Since $a\not \equiv l \: (\mbox{mod } 2)$, $h=\frac{a-1-l}{2}+2$. 
 Follow the same
 procedure as for the case 1  except
 that this time there are $\frac{a-1-l}{2}$ remaining agents to be
 redistributed once $S_l'$ has been formed.  Therefore, when we
 redistribute all these agents among the coalitions $S_{1},\ldots
 S_{l-1}$, we get all coalitions that were found in part 1, along with
 coalitions of size $h-1$. ${}_\Box$
\end{enumerate}
\end{proofnobox}
From Lemma~\ref{lem-pairing}, it follows that after searching level
$l$ with Algorithm~\ref{al:coal_search}, we cannot have seen two
coalitions of $h$ members together in the same coalition structure.
 
\begin{theorem}
\label{thm-bound}
After searching level $l$ with
Algorithm~\ref{al:coal_search}, the bound $k(n)$ is $\lceil
\frac{a}{h}\rceil$ if $a\equiv h-1 \: (\mbox{mod }h)$ and $a\equiv l
\:(\mbox{mod } 2)$.  Otherwise the bound is $\lfloor
\frac{a}{h}\rfloor $.
\end{theorem}

\begin{proof}
Case 1.  Assume $a\equiv l \:(\mbox{mod }2)$  and  $a\equiv h-1 \:(\mbox{mod } h-1)$. 
 Let $\alpha $ be an
 assignment of coalition values which give the worst case. For
 any other assignment of coalition values,  $\beta $,  the 
 inequality 
$k(n)=\frac{V_{\alpha}(CS^{*})}{V_{\alpha }(CS_l)}\geq \frac{V_{\beta}(CS^{*})}{V_{\beta}(CS_l)}$  holds.
  Since $CS^{*}$ is the best coalition structure under $\alpha $, we can
assume that $V_{S}=0$ for all coalitions $S\not \in CS^{*}$ without
decreasing the ratio $\frac{V_{\alpha }(CS^{*})}{V_{\alpha
}(CS_l)}$.  Also, no two coalitions $S,S'\in CS^{*}$ can appear
together if $v_{S}+v_{S'}>\mbox{max}\{v_{S^{``}}\} $ for $S^{``}\in CS^{*}$, 
since otherwise we could decrease the ratio $k(n)$.
Therefore $V_{\alpha }(CS_l)=\mbox{max}\{v_{S}\}$ for $S\in CS^{*}$ Call this value 
$v^{*}$.
We can derive an equivalent worst case, $\alpha'$, from $\alpha$ as
follows:
\begin{enumerate}
\item Find a coalition structure $CS'$ with $\lfloor \frac{a}{h}\rfloor $ coalitions
of size $h$ and one coalition of size $h-1$.
\item Define $\overline{v}=\frac{V_{\alpha}(CS^{*})}{\lfloor \frac{a}{h}\rfloor +1}$.
\item Assign a value $v'_{S}=\overline{v}$ to each coalition in $CS'$.
\end{enumerate}
Clearly $V_{\alpha} (CS^{*})=V_{\alpha'}(CS')$.  From
Lemma~\ref{lem-pairing} we know that no two coalitions in $CS'$
have been seen together.  The best value of a coalition structure
seen during the search is $V_{\alpha'}(CS_l)=\overline{v}$.
Therefore the following inequalities hold;
\begin{displaymath}
 V_{\alpha'}(CS')=(\lfloor \frac{a}{h}\rfloor +1) \overline{v}=(\lfloor \frac{a}{h}\rfloor +1) V_{\alpha'}(CS_l) 
\end{displaymath} 
\begin{displaymath}
(\lfloor \frac{a}{h}\rfloor +1) V_{\alpha'}(CS_l) \leq (\lfloor \frac{a}{h}\rfloor +1) v^{*}\leq (\lfloor \frac{a}{h}\rfloor +1) V_{\alpha}(CS_l). 
\end{displaymath}
Since $V_{\alpha} (CS^{*})=V_{\alpha'}(CS')$ and $V_{\alpha
'(CS_l)}\leq V_{\alpha }(CS_l)$,
\begin{displaymath}
 k(n)=\frac{V_{\alpha }(CS^{*})}{V_{\alpha }(CS_l)}\leq \frac{ V_{\alpha'}(CS^{'
})}{V_{\alpha'(CS_l)}} = \lfloor \frac{a}{h}\rfloor +1 = \lceil
\frac{a}{h}\rceil. 
\end{displaymath}
 Therefore the bound is $\lceil \frac{a}{h}\rceil$.\newline
Case 2.  This is a similar argument as in Case 1, except that the
assignment of values to the coalitions in the equivalent worst case
coalition structure is different.  Define $\alpha $ as before and let $CS^+$
 be a coalition structure with $\lfloor \frac{a}{h}\rfloor $ coalitions
of size $h$ and one possible remainder coalition of size less than $h$.
 Define $\overline{v}=\frac{V_{\alpha}(CS^{*})}{\lfloor \frac{a}{h} \rfloor }$ and 
 assign  value $v^+_{S}=\overline{v}$ if $|S|=h$ and $S\in CS^+$,
 otherwise let $v_{S}=0$ for all other coalitions including the remainder coalition in $CS^+$.
 Thus the best coalition seen has
value $V_{\alpha ^+}(CS_l)=\overline{v_{S}}$ and we have the following 
inequalities:
\begin{displaymath}
 V_{\alpha ^+}(CS^+)=(\lfloor \frac{a}{h}\rfloor ) \overline{v}=(\lfloor \frac{a}{h}\rfloor ) V_{\alpha^+}(CS_l) 
\end{displaymath}
\begin{displaymath} 
(\lfloor \frac{a}{h}\rfloor ) V_{\alpha^+}(CS_l) \leq (\lfloor \frac{a}{h}\rfloor ) v^{*}\leq (\lfloor \frac{a}{h}\rfloor ) V_{\alpha}(CS_l). 
\end{displaymath}
 Therefore the bound $k(n)=\lfloor \frac{a}{h}\rfloor $.
\end{proof}
\begin{theorem}
\label{thm-tight}
The bound in Theorem~\ref{thm-bound} is tight.
\end{theorem}
 \begin{proof}  Case 1: Assume $a\equiv l \:(\mbox{mod }2)$ and $a\equiv h-1 \:
 (\mbox{mod } h)$.  The bound is
 $\lceil \frac{a}{h} \rceil$.  Assume you have the coalition structure
 $CS'$ from Theorem~\ref{thm-bound}.  Assign value 1 to each
   coalition $S\in CS'$ and assign value 0 to all other coalitions.
   Then $V(CS')=\lceil \frac{a}{h}\rceil $.   Since
  ( Lemma~\ref{lem-pairing}) no two of the coalitions in $CS'$  have
  ever appeared in the same coalition structure, $V(CS_l)=1$.
   Therefore
 $ \frac{V(CS')}{V(CS_l)} = \frac{\lceil \frac{a}{h} \rceil }{1}=\lceil \frac{a}{h}\rceil $
   and the bound is tight.\newline
   Case 2: Assume  $a\not \equiv l
  \:(\mbox{mod }2)$ or $a\not \equiv h-1 \: (\mbox{mod } h)$.
  The bound is $\lfloor \frac{a}{h}\rfloor$.
   Assign value 1 to each coalition $S\in
  CS^+$ from Theorem~\ref{thm-bound} and assign value 0 to all other coalitions.  Then
  $V(CS^+)=\lceil \frac{a}{h}\rceil $ and  $V(CS_l)=1$.
  Therefore
 $ \frac{V(CS^+)}{V(CS_l)} = \frac{\lfloor \frac{a}{h} \rfloor }{1}=\lfloor \frac{a}{h}\rfloor $
 and the bound is tight.
 \end{proof}

\begin{figure}[hbt]
\vspace{-0.35in}
\epsfxsize=2.3in
\hspace*{\fill}
\hspace{-0.1in}
\rotate[r]{
\epsffile{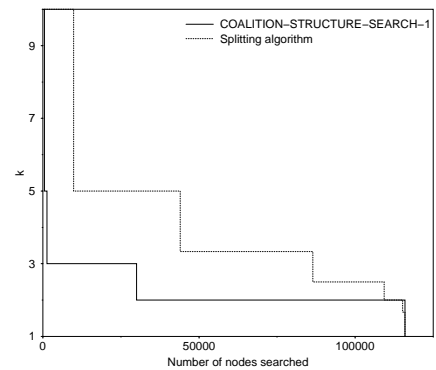}
}
\hspace*{\fill}
\vspace{-0.15in}
\caption[Ratio bound as a function of search size in a 10-agent game.]{{\it Ratio
bound $k$ as a function of search size in a 10-agent game.}}
\label{fi:k_in_n}
\end{figure}

As we have shown in the previous section, before $2^{a-1}$ nodes have
been searched, no bound can be established, and at $n=2^{a-1}$ the
bound $k=a$.  The surprising fact is that by seeing just one
additional node ($n=2^{a-1} +1$), i.e. the top node, the bound drops
in half ($k=\frac{a}{2}$).  Then, to drop $k$ to about $\frac{a}{3}$,
two more levels need to be searched.  Roughly speaking, the divisor in
the bound increases by one every time two more levels are searched.
So, the anytime phase (step 2) of Algorithm~\ref{al:coal_search} has
the desirable feature that the bound drops rapidly early on, and there
are overall diminishing returns to further search,
Figure~\ref{fi:k_in_n}.


\subsection{Comparison to other algorithms}

All previous coalition structure generation algorithms for general
CFGs~\cite{Shehory96:Kernel,Ketchpel:aaai94}---that we know of---fail
to establish any worst case bound because they search fewer than
$2^{a-1}$ coalition structures.  Therefore, we compare our
Algorithm~\ref{al:coal_search} to two other obvious candidates:
\begin{itemize}
\item {\bf Merging algorithm}, i.e. breadth first search from the top
of the coalition structure graph.  This algorithm cannot establish any
bound before it has searched the entire graph.  This is because, to
establish a bound, the algorithm needs to see every coalition, and the
grand coalition only occurs in the bottom node.  Visiting the grand
coalition as a special case would not help much since at least part of
level 2 needs to be searched as well: coalitions of size $a-2$ only
occur there.


\item {\bf Splitting algorithm}, i.e. breadth first search from the
bottom of the graph.  This is identical to
Algorithm~\ref{al:coal_search} up to the point where $2^{a-1}$ nodes
have been searched, and a bound $k=a$ has been established.  After
that, the splitting algorithm reduces the bound much slower than
Algorithm~\ref{al:coal_search}.  This can be shown by constructing bad
cases for the splitting algorithm: the worst case may be even worse.
To construct a bad case, set $v_S = 1$ if $|S|=1$, and $v_S = 0$
otherwise.  Now, $CS^* = \{\{1\}, ..., \{a\}\}$, $V(CS^*) = a$, and
$V(CS^*_N) = l-1$, where $l$ is the level that the algorithm has
completed (because the number of unit coalitions in a $CS$ never
exceeds $l-1$).  So, $\frac{V(CS^*)}{V(CS^*_N)} =
\frac{a}{l-1}$,\footnote{The only exception comes when the algorithm
completes the last (top) level, i.e $l=a$.  Then
$\frac{V(CS^*)}{V(CS^*_N)} =1$.}  Figure~\ref{fi:k_in_n}.  In other
words the divisor drops by one every time a level is searched.
However, the levels that this algorithm searches first have many more
nodes than the levels that Algorithm~\ref{al:coal_search} searches
first.
\end{itemize}

\subsection{Variants of the problem}

In general, one would want to construct an anytime algorithm that
establishes a lower $k$ for any amount of search $n$, compared to any
other anytime algorithm.  However, such an algorithm might not exist.
It is conceivable that the search which establishes the minimal $k$
while searching $n'$ nodes ($n' > n$) does not include all nodes of
the search which establishes the minimal $k$ while searching $n$
nodes.  This hypothesis is supported by the fact that the curves in
Figure~\ref{fi:k_in_n} cross in the end.  However, this is not
conclusive because Algorithm~\ref{al:coal_search} might not be the
optimal anytime algorithm, and because the bad cases for the splitting
algorithm were not shown to be worst cases.

If it turns out that no anytime algorithm is best for all $n$, one
could use information (e.g. exact, probabilistic, or bounds) about the
termination time to construct a {\em design-to-time algorithm} which
establishes the lowest possible $k$ for the specified amount of
search.

In this paper we have discussed algorithms that have an {\em off-line
search control} policy, i.e. the nodes to be searched have to be
selected without using information accrued from the search so far.
With {\em on-line search control}, one could perhaps establish a lower
$k$ with less search because the search can be redirected based on the
values observed in the nodes so far.  With on-line search control, it
might make a difference whether the search observes only values of
coalition structures, $V(CS)$, or values of individual coalitions,
$v_S$, in those structures.  The latter gives more information.

None of these variants (anytime vs. design-to-time, and off-line
vs. on-line search control) would affect our results that searching
the bottom two levels of the coalition structure graph is the unique
minimal way to establish a worst case bound, and that the bound is
tight.  However, the results on searching further might vary in these
different settings. This is a focus of our future research.



\section{Distributing coalition structure search among insincere agents}

%


This section discusses the distribution of coalition structure search
across agents (because the search can be done more efficiently in
parallel, and the agents will share the burden of computation) and the
methods of motivating self-interested agents to actually follow the
desired search method.  Self-interested agents prefer greater
personal payoffs, so they will search for coalition structures that
maximize personal payoffs, ignoring $k$.  In order to motivate such
agents to follow a particular search that leads to desirable social
outcomes (e.g. a search that guarantees a desirable worst case bound
$k$), the interaction protocol has to be carefully designed.  It is
also necessary to take into account that an agent's preference between
$CS$s depends on the way in which $V(CS)$ is distributed among the
agents.  Classical game theoretic $CS$ selection and payoff division
methods are not viable (unless modified) in our setting since they
require knowledge of every $CS \in M$.  This is because, according to
those solution concepts, an agent can justifiably claim more than
others receive from $V(CS)$ by objecting to $CS$ (both to the
structure and to the payoff distribution).  A justified objection (as
defined classically e.g. in~\cite{Kahan84:Theories}) depends on all
possible $CS$s. Thus it uses information beyond the region of the
search space that any nonexhaustive algorithm should search.  The
protocol designer cannot prohibit access to information that may
support such objection because the agents can locally decide what to
search.  However, the protocol designer can forbid objections and make
additional search unbeneficial, as we demonstrate below.\footnote{The
protocol designer cannot prevent agents from opting out, but such
agents receive null excess payoffs since they do not collude with
anyone.  That is, for $|S|=1$, the payoff to the agent is equal to its
coalition value $v_S$.  This assumes that agents do not recollude
outside the protocol, but such considerations are outside of protocol
design.} The distributed search consists of the following stages:
\begin{enumerate}

\item
{\bf Deciding what part of the coalition structure graph to search:} \
This decision can be made in advance (outside the distributed search
mechanism), or be dictated by a central authority, or by a randomly
chosen agent\footnotemark[9], or be decided using some form of
negotiation.  The earlier results in this paper give prescriptions
about which part to search.  For example, the agents can decide to use
Algorithm~\ref{al:coal_search}.


\item
{\bf Partitioning the search space among agents:} \ Each agent is
assigned some part of the coalition structure graph to search.  The
enforcement mechanism, presented later, will motivate the agents to
search exactly what they are assigned, no matter how unfairly the
assignment is done.  One way of achieving {\em ex ante} fairness is to
randomly allocate the set search space portions to the agents.  In
this way, each agent searches equally on an expected value basis,
although {\em ex post}, some may search more than others.\footnote{The
randomization can be done without a trusted third party by using a
distributed nonmanipulable protocol for randomly permuting 
agents~\cite{Zlotkin:aaai94}.  Distributed randomization is also
discussed in~\cite{Linial92:Games}.}  The fairest option is to
distribute the space so that each agent gets an equal share.



\item
{\bf Actual search:} \ Each agent searches part of the search space.
The enforcement mechanism guarantees that each agent is motivated to
search exactly the part of the space that was assigned to that agent.
Each agent, having completed the search, tells the others which $CS$
maximized $V(CS)$ in its search space.

\item
{\bf Enforcement of the protocol:} \ One agent, $i$, and one search
space of an agent $j$, $j\neq i$, will be selected
randomly.\footnotemark[9] Agent $i$ will re-search the search space of
$j$ to verify that $j$ has performed its search as required.  Agent
$j$ gets caught of mis-searching (or misrepresenting) if $i$ finds a
better $CS$ in $j$'s space than $j$ reported (or $i$ sees that the
$CS$ that $j$ reported does not belong to $j$'s space at all).  If $j$
gets caught, it has to pay a penalty $P$.  To motivate $i$ to conduct
this additional search, we make $i$ the claimant of $P$.
%
%
There is no pure strategy Nash equilibrium in this
protocol.\footnote{See~\cite{Mas-Colell95:Microeconomic} for a
definition of Nash equilibrium.}  If $i$ searches and the penalty is
high enough, then $j$ is motivated to search sincerely.  However, $i$
is not motivated to search since it cannot receive $P$.  Instead,
there will be a mixed strategy equilibrium where $i$ and $j$ search
truthfully with some probabilities.  By increasing $P$, the
probability that $j$ searches can be made arbitrarily close to one.
The probability that $i$ searches approaches zero, which minimizes
enforcement overhead.\footnote{Agent $j$ will try to trade off the
cost of search against the risk of getting caught, and could decide
that the risk is worth taking.  This problem can be minimized by
choosing a high enough $P$.}



\item {\bf Additional search:} \ The previous steps of this
distributed mechanism can be repeated if more time to search remains.
For example, the agents could first do step 1 of
Algorithm~\ref{al:coal_search}.  Then, they could repeatedly search
more and more as time allows, again using the distributed method.

\item
{\bf Payoff division:} \ Many alternative methods for payoff division
among agents could be used here.  The only concern is that the
division of $V(CS)$ may affect what $CS$ an agent wants to report as a
result of its search since different $CS$s may give the agent
different payoffs (depending on the payoff division scheme).  However,
by making $P$ high enough compared to $V(CS)$s, this consideration can
be made negligible compared to the risk of getting caught.

\end{enumerate}

\section{Related research on computational coalition formation}
\label{se:related_research}

Coalition formation has been widely studied in game
theory~\cite{Kahan84:Theories,Bernheim87:Coalition-Proof,Aumann59:Acceptable}.
They address the question of how to divide $V(CS^*)$ among agents so
as to achieve stability of the payoff configuration.  Some also
address coalition structure generation.  However, most of that work
has not taken into account the computational limitations involved.
This section reviews some of the work that has.

\cite{Deb96:Nakamura} bound the maximal number of payoff
configurations that must be searched to guarantee stability.
Unlike our work, they neither address a bound on solution quality nor
provide methods for coalition structure generation.

~\cite{Ketchpel:aaai94} presents a coalition formation method which
addresses coalition structure generation as well as payoff
distribution.  These are handled simultaneously.  His algorithm uses
cubic time in the number of agents, but guarantees neither a bound
from optimum nor stability of the coalition structure.  There is no
mechanism for motivating self-interested agents to follow his
algorithm.


~\cite{Shehory96:Kernel} analyze coalition formation among
self-interested agents with perfect information in CFGs.  Their
protocol guarantees that if agents follow it (nothing necessarily
motivates them to do so), a certain stability criterion (K-stability)
is met.  Their other protocol guarantees a weaker form of stability
(polynomial K-stability), but only requires searching a polynomial
number of coalition structures.  Their algorithm is an anytime
algorithm, but does not guarantee a bound from optimum.

~\cite{Shehory95:Task} also present an algorithm for coalition
structure generation among cooperative agents.  The complexity of the
problem is reduced by limiting the number of agents per coalition. The
greedy algorithm guarantees that the solution is within a loose ratio
bound from the best solution that is possible {\em given the limit on
the number of agents}.  However, this benchmark can, itself, be
arbitrarily far from optimum.  On the other hand, our work computes
the bound based on the actual optimum.  Our result that no algorithm
can establish a bound while searching less than $2^{a-1}$ nodes does
not apply to their setting because they are not solving general CFGs.
Instead, they address a more specialized setting where the $v_S$
values have special structure.  In such settings it may be possible to
establish a worst case bound with less search than in general CFGs.

~\cite{Sandholm97:Coalition_incl95} study coalition formation with a focus on
the optimization activity: how do computational limitations affect
which coalition structure should form, and whether that structure is
stable?  That work used a normative model of bounded rationality based
on the agents' algorithms' performance profiles and the unit cost of
computation.  All coalition structures were enumerated because the
number of agents was relatively small, but it was not assumed that
they could be evaluated exactly because the optimization problems
could not be solved exactly due to intractability.  The methods of
this paper can be combined with their work if the performance profiles
are deterministic.  In such cases, the $v_S$ values represent the
value of each coalition, given that that coalition would strike the
optimal tradeoff between quality of the optimization solution and the
cost of that computation.  Our algorithm can be used to search for a
coalition structure, and only afterwards would the coalitions in the
chosen coalition structure actually attack their optimization
problems.  If the performance profiles include uncertainty, this
separation of coalition structure generation and optimization does not
work e.g. because an agent may want to redecide its membership if its
original coalition receives a worse optimization solution than
expected.

\section{Conclusions and future research}
\label{se:conclusion_future}

Coalition formation is a key topic in multiagent systems.  One would
prefer a coalition structure that maximizes the sum of the values of
the coalitions, but often the number of coalition structures is too
large to allow exhaustive search for the optimal one.  This paper
focused on establishing a worst case bound on the quality of the
coalition structure while only searching a small portion of the
coalition structures.

We showed that none of the prior coalition structure generation
algorithms for general CFGs can establish any bound because they
search fewer nodes than a threshold that we showed necessary for
establishing a bound.  We presented an algorithm that establishes a
tight bound within this minimal amount of search, and showed that any
other algorithm would have to search strictly more.  The fraction of
nodes needed to be searched approaches zero as the number of agents
grows.

If additional time remains, our anytime algorithm searches further,
and establishes a progressively lower tight bound.  Surprisingly, just
searching one more node drops the bound in half.  As desired, our
algorithm lowers the bound rapidly early on, and exhibits diminishing
returns to computation.  It also drastically outperforms its obvious
contenders: the merging algorithm and the splitting algorithm.
Finally, we showed how to distribute the desired search across
self-interested manipulative agents.

Our results can also be used as prescriptions for designing
negotiation protocols for coalition structure generation.  The agents
should not start with everyone operating separately---as one would do
intuitively.  Instead, they should start from the grand coalition, and
consider different ways of splitting off exactly one coalition.  After
that, they should try everyone operating separately, and continue from
there by considering mergers of two coalitions at a time.

Future research includes studying design-to-time algorithms and
on-line search control policies for coalition structure generation.
We are also analyzing the interplay of dynamic coalition formation and
belief revision among bounded-rational
agents~\cite{Tohme97:Coalition}. 
%
%
The long term goal is to construct normative methods that reduce the
complexity---in the number of agents and in the size of each
coalition's optimization problem---for coalition structure generation,
optimization and payoff division.

{\footnotesize
\bibliographystyle{$HOME/sty/aaai98}
\bibliography{/home/cs/group/mas/refs/dairefs,temprefs}
}
\end{document}